\documentclass[aps,prb,twocolumn,epsfig,floats,superscriptaddress]{revtex4}
\usepackage{amsmath}
\usepackage{color}
\usepackage{graphicx}
\usepackage{amsfonts}
\usepackage{mathrsfs}
\usepackage{amsmath}
\usepackage{amssymb}
\usepackage{times}
\usepackage{hyperref}
\hypersetup{
  colorlinks=true,
  citecolor=blue,
  linkcolor=blue,
  urlcolor=blue}

\usepackage[version=4]{mhchem}

\begin{document}

\title{\textbf{Phases and collective modes of bosons in a triangular lattice at finite temperature: A cluster mean field study}}

\author{M. Malakar}
\affiliation{Indian Institute of Science Education and Research Kolkata, Mohanpur, Nadia 741246, India}

\author{S. Ray}
\affiliation{Department of Chemistry, Ben-Gurion University of the Negev, Beer-Sheva 84105, Israel}

\author{S. Sinha}
\affiliation{Indian Institute of Science Education and Research Kolkata, Mohanpur, Nadia 741246, India}

\author{D. Angom}
\affiliation{Physical Research Laboratory, Ahmedabad 380009, Gujarat, India}

%\date{\today}
\begin{abstract}

Motivated by the realization of Bose-Einstein condensates (BEC) in non-cubic lattices, in this
work we study the phases and collective excitation of bosons with nearest neighbor interaction in a
triangular lattice at finite temperature, using mean field (MF) and cluster mean field (CMF) theory.
We compute the finite temperature phase diagram both for hardcore and softcore bosons, as well analyze the effect of correlation arising due to lattice frustration and interaction systematically using CMF method. A semi-analytic estimate of the transition temperatures between different phases are derived within the framework of MF Landau theory, particularly for hardcore bosons. Apart from the usual phases such as density waves (DW) and superfluid (SF), we also characterize different supersolids (SS). These phases and their transitions at finite temperature are identified from the collective modes. The low lying excitations, particularly Goldstone and Higgs modes of the supersolid can be detected in the ongoing cold atom experiments.

\end{abstract}

%\date{}
\maketitle

%%%%%%%%%%%%%%%%%%%%%%%%%%%%%%%%%%%%%%%%%%%%%%%%%%%%%%%%%%%%%%%%%%%%%%%%%%%%%%%%%%%%%
\section{Introduction}
\label{intro}

Frustrated lattice systems are one of the most active research areas of condensed matter physics which has led to the observation of various exotic phases of matter \cite{Lacroix11} such as spin liquids \cite{Balents10}, spin ice state in pyrochlore material \cite{Harris97,Bramwell01}, as well magnetic phases and phase transition \cite{Wannier50,Fennel09}. In recent experiments anti-ferromagnetic spin models in triangular lattice have been realized in the complex compounds like \ce{Ba3CoSb2O9}, and its magnetization process, specific heat as well as the collective excitation are also measured \cite{Shirata12,Susuki13,Shen16}. Realization of Bose-Einstein condensates (BEC) in non-cubic lattice geometries e.g. experimental demonstration of superfluid-Mott insulator (SF-MI) transition in triangular and hexagonal optical lattices \cite{Becker10} have further opened up the possibility to explore the competition between interaction and geometric frustration. Ultracold bosonic atoms trapped in a triangular optical lattice has further paved the way to study different magnetic phases of frustrated classical spin models \cite{Struck11} and in the presence of synthetic gauge field \cite{Struck13}. The existence of different types of supersolid phases, and their melting driven by either quantum fluctuation or thermal fluctuation in triangular lattices have been theoretically investigated \cite{Murthy97,Wessel05,Damle05,Balents05,Prokofev05}.

Supersolid is a state of matter where particles are organized in a crystalline order, and show a dissipation-less superflow \cite{Lifshitz69,Leggett70,Prokofev07}. 
Such a phase of matter has been predicted in a number of theoretical studies since past many years particularly in bosonic systems with long range interaction in optical lattices \cite{Fisher72,Zimanyi95,Sarma05,Sinha05,SenguptaP05,Nath19}, Josephson junction arrays \cite{Fazio95,Stroud95}, in a Bose-Fermi mixture \cite{Blatter03,Hofstetter08,Sinha09} and so on. 
As a result of experimental progress in ultracold atomic systems, quantum gases with dipolar interaction \cite{Kurizki02,Santos02}, spin-orbit coupled condensate \cite{Stringari12} and Rydberg gases \cite{Pohl10,Saha14,Hofstetter18} have become promising candidates to search for supersolid phase.
In recent cold-atom experiments supersolid has been observed in spin-orbit coupled Bose-Einstein condensates (BEC) \cite{Ketterle17}, BEC in optical lattice \cite{Esslinger16} and coupled to optical cavity \cite{Esslinger17}, trapped dipolar BEC of Erbium (Er) and Dysprosium (Dy) atoms \cite{Ferlaino19,Pfau19,Stringari19}. Apart from the density modulation revealing the crystalline order, the signature of $U(1)$ symmetry breaking has also been confirmed experimentally from the low energy collective excitation such as Goldstone and Higgs modes \cite{Esslinger17,Ferlaino19,Pfau19,Stringari19}.

On the other hand a stable supersolid formation due to the competition between particle interaction and frustration in a triangular lattice has been predicted in a number of theoretical studies \cite{Murthy97,Wessel05,Damle05,Balents05,Prokofev05,Gan07,Pollmann09,Pollet10,Yamamoto12}. A supersolid phase of Rydberg excited atoms in a triangular lattice has also been predicted \cite{Hofstetter19}. At finite temperature the equilibrium phases of hardcore bosons in a triangular lattice have been studied \cite{Prokofev05}. Superfluid to Mott insulator transition at finite temperatures in cubic lattice has also been investigated theoretically \cite{Prokofev08,Rozek15,Trivedi11,Pinaki20}. However, incorporating the correlation systematically to understand the interplay between lattice frustration and interaction at finite temperature, moreover to characterize these phases from their collective excitation is beyond the scope of these studies. 
Motivated by the recent experiments, in this work we primarily chart out the phases of bosons in a triangular lattice at finite temperature using the {\it cluster mean field} (CMF) technique, and compute the collective excitation for both the following cases, one with onsite hardcore repulsion i.e. $U \rightarrow \infty$, and another with finite $U$ which is a more realistic scenario.
We supplement semi-analytical results obtained from the Landau-Ginzburg theory which qualitatively captures the numerically observed phases of hardcore bosons at finite temperature. 
The transition between these phases is captured from the collective modes which can be detected in the cold atom experiments.

The paper is organized as follows. In Sec.\ \ref{model} we describe the Bose-Hubbard model on a triangular lattice and demonstrate the {\it cluster mean field} method extended to finite temperature. In Sec.\ \ref{MF_HCB} we provide the {\it mean field} phase diagram of hardcore bosons followed by the semi-analytic estimate of the transition temperature from Landau-Ginzburg theory. The linear stability of these phases are analyzed and their collective excitation are computed in Sec.\ \ref{excitation_hcb}. The effect of correlation is discussed using {\it cluster mean field} in Sec.\ \ref{CMF_HCB}, and the results are compared with the existing Quantum Monte Carlo (QMC) studies. In Sec.\ \ref{scb} we discuss the {\it zero} and {\it finite} temperature phases of bosons with finite onsite repulsion, and compute their collective modes. Finally we summarize our work and conclude in Sec.\ \ref{conclu}.    

\section{Model and the method}
\label{model}     

The Bose-Hubbard model with nearest neighbor interaction in the grand canonical ensemble can be described in general by the Hamiltonian,
\begin{eqnarray}
\hat{H} &=& -t\sum_{\langle i,j\rangle} (\hat{a}_i^{\dagger}\hat{a}_j + \text{h.c.}) -\mu \sum_{i} \hat{n}_i + V \sum_{\langle i,j\rangle} \hat{n}_i \hat{n}_j \nonumber \\
&+& \frac{U}{2} \sum_{i} \hat{n}_i(\hat{n}_i-1)    
\label{Ham}
\end{eqnarray} 
where, $\hat{a}_i^{\dagger} (\hat{a}_i)$ are the bosonic creation (annihilation) operator at the $i$th site, $\hat{n}_i$ represents the local number operator, $t$ and $V$ are the hopping amplitude and interaction strength respectively between the nearest neighbor sites of the triangular lattice denoted by $\langle i,j\rangle$, $U$ is the onsite interaction, and $\mu$ is the chemical potential. In what follows we set $\hbar = 1$, Boltzmann constant $k_B = 1$, and measure all the energies in the unit of interaction strength $V$ unless it is otherwise mentioned.  

\begin{figure}[ht]
\centering
\includegraphics[width=0.7\columnwidth]{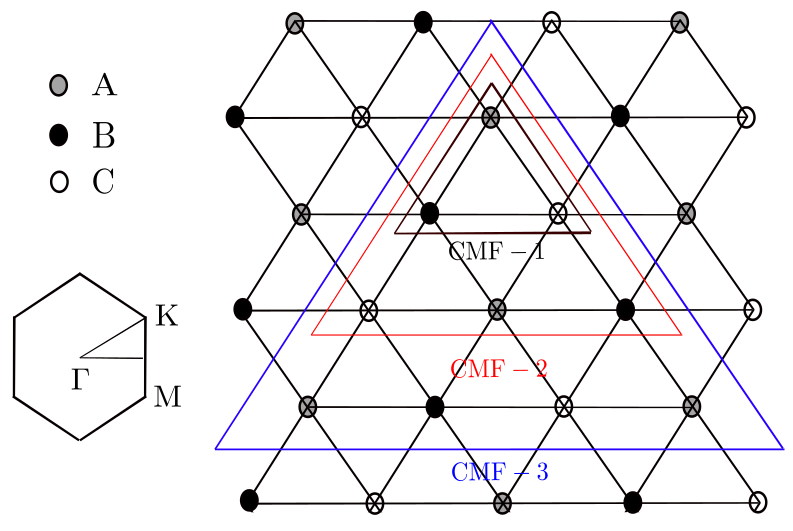}
\caption{Schematic representation of the triangular lattice and its three sublattice structure. The reduced hexagonal Brillouin zone is shown in the left. The different sized clusters used in {\it cluster mean field} are shown by the enclosing triangles.}
\label{cmf_triangle}
\end{figure}

We focus on the equilibrium phases of the above model at a finite temperature $T$ using {\it cluster mean field} (CMF) method. Such a method has been used previously to study zero temperature phases and non-equilibrium dynamics of bosons in an optical lattice \cite{Yamamoto12,Demler12,Mishra14,Fazio16}. We extend this to finite temperature and study the effect of correlation by considering a cluster $\mathcal{C}$ of different sizes in a triangular lattice as illustrated in Fig.\ \ref{cmf_triangle}. While all the correlations are considered exactly within the cluster $\mathcal{C}$ of a given size using exact diagonalization, the interaction and hopping between edge sites of $\mathcal{C}$ and its neighboring sites outside $\mathcal{C}$ are treated at the mean field level. More generically the composite Hamiltonian can be written as,
\begin{equation}
\hat{H} = \hat{H}_{\mathcal{C}} + \hat{H}_{\mathrm{MF}}
\end{equation}    
where, $\hat{H}_{\mathcal{C}}$ describes the bosons within the cluster and the {\it mean field} term corresponding to the edge sites can be written as,
\begin{eqnarray}
\hat{H}_{\mathrm{MF}} &=& \sum_{i \in \text{edge sites}} \hat{H}_{\mathrm{MF}}^{i} \nonumber \\
\hat{H}_{\mathrm{MF}}^{i} &=& \sum_{\langle i,j\rangle, ~ j \not\in \mathcal{C}} \left[-J (\alpha_j \hat{a}_i^{\dagger} + \alpha_j^{*} \hat{a}_i) + V n_j \hat{n}_i \right]
\end{eqnarray}
Exploiting the sublattice symmetry throughout the lattice, the {\it mean field} values $\alpha_j$ and $n_j$ can be obtained from $\langle \hat{a}_{\text{A,B,C}}\rangle$ and $\langle \hat{n}_{\text{A,B,C}}\rangle$ respectively where the sub-lattices A, B and C belong to the cluster $\mathcal{C}$. Note that in case of zero temperature the average of observable is denoted by $\langle .\rangle \equiv \langle \psi |.|\psi \rangle$, $|\psi \rangle$ being the ground state of $\hat{H}$. At a finite temperature this amounts to thermal average, $\langle .\rangle \equiv \text{Tr} (\hat{\rho}_T ~.)$, where the thermal density matrix is given by, 
\begin{equation}
\hat{\rho}_T = \frac{e^{-\beta \hat{H}}}{Z}, \quad Z = \text{Tr}(e^{-\beta \hat{H}}), \quad \beta = 1/T
\end{equation}
Solving the mean field self consistency we finally obtain the equilibrium density matrix $\hat{\rho}_T$ at a finite temperature $T$ with converged free energy $F = E - TS$, where $E = \langle \hat{H}\rangle$ and von Neumann entropy $S = -\text{Tr} (\hat{\rho}_T \log \hat{\rho}_T)$. Different physical quantities at finite temperature can be obtained by averaging over the cluster using the density matrix $\hat{\rho}_T$.

\section{Phases and collective excitation of hardcore boson}
\label{hcb}

In this section we discuss the phases and collective modes of bosons in a triangular lattice with hardcore repulsion i.e. $U \rightarrow \infty$ allowing not more than one boson per site. First we present the phase diagram using standard {\it mean field} calculation and subsequently show how the results get improved by introducing clusters and thereby incorporating correlations in the system.

\subsection{Mean field phase diagram} 
\label{MF_HCB}

\begin{figure}[t]
\centering
\includegraphics[clip=true, width=1\columnwidth]{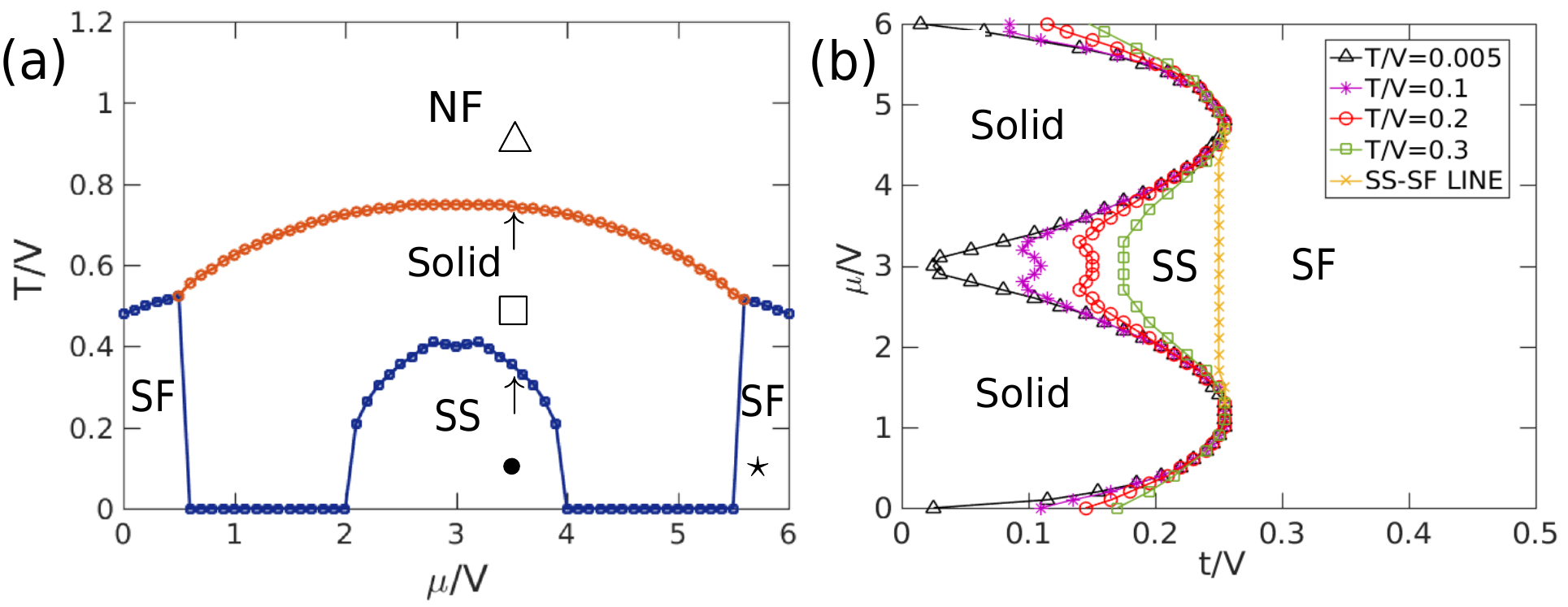}
\caption{{\it Finite temperature mean field phase diagram of hardcore bosons in triangular lattice:} (a) in $\mu$ vs $T$ plane for $t = 0.2$, and (b) in $\mu$ vs $t$ plane for different temperatures mentioned in the inset. Collective excitation at the points marked by bullets in (a) are shown in Fig.\ \ref{Excitation_hcb_T}. Vanishing of respective order parameters and excitation energy gap at SS-solid and solid-NF boundary (marked by $\uparrow$) are depicted in Fig.\ \ref{Egap_OP}.}
\label{PD_MF}
\end{figure}

We consider one unit cell consisting of three sub-lattices A, B and C which are decoupled at the mean field level. Therefore the total density matrix of a unit cell in a triangular lattice can be written as,
\begin{equation}
\hat{\rho}_T = \prod_i \hat{\rho}_i, \quad 
\hat{\rho}_i = \left[ \begin{array}{cc}
\frac{1}{2}(1-m_i) & \alpha_i^* \\
\alpha_i &  \frac{1}{2}(1+m_i)\end{array}\right]
\end{equation}  
where, $\hat{\rho}_i$, $i \in \{A,B,C\}$ represents a thermal density matrix satisfying $\text{Tr}[\hat{\rho}_i] = 1$. It can be noted that by definition local SF order parameter is, $\langle \hat{a}_i\rangle = \alpha_i$, and $m_i$ is related to the local density as, $\langle \hat{n}_i\rangle = (1+m_i)/2$. 
Thus the corresponding {\it mean field} free energy at temperature $T$ for a given site can be written as,
\begin{eqnarray}
F_{\text{MF}}^i &=& -\frac{3t}{2}\left(\alpha_i^*\sum_{\bar{i}\neq i}\alpha_{\bar{i}} + \alpha_i\sum_{\bar{i}\neq i}\alpha_{\bar{i}}^*\right) - \frac{\mu}{2}\left(1+m_i\right) \nonumber \\
&+& \frac{3V}{8}\left(1+m_i\right)\sum_{\bar{i}\neq i}\left(1+m_{\bar{i}}\right) - TS_i
\label{free_en}
\end{eqnarray}
where, $i$ and $\bar{i}$ denote nearest neighbor sites $\in \{A,B,C\}$, and the von Neumann entropy $S_i$ is given by,
\begin{equation}
S_i = -\sum_{\sigma = +,-} \lambda_{\sigma}^i \log \lambda_{\sigma}^i, \quad \lambda_{\pm} = \frac{1 \pm \sqrt{m_i^2+4|\alpha_i|^2}}{2}
\end{equation}
Minimizing the average free energy $F=\sum_{i} F_{\text{MF}}^i/3$ with respect to the order parameters $\alpha_i$'s and $m_i$'s we obtain the phase diagram in the $\mu$ vs $T$ plane for a fixed tunneling amplitude $t$ as shown in Fig. \ref{PD_MF}a. At low temperature, apart from homogeneous superfluid two types of density waves exist, which at $T=0$ have a sublattice density structure $(1,0,0)$ and $(1,1,0)$ for $\mu<3$ and $\mu>3$ respectively; as $t$ is increased they melt to form supersolids namely SSA and SSB for $\mu \lessgtr 3$ \cite{Murthy97,Wessel05}. At a finite $T$, these two solid lobes and the supersolid are shown in Fig.\ \ref{PD_MF}b. It can be noted that as $T \rightarrow 0$, the phase boundaries agree with the {\it zero} temperature {\it mean field} phase diagram \cite{Murthy97}. With increasing temperature we observe transition from (i) superfluid (SF) to normal fluid (NF), (ii) density wave (DW)/solid to NF, and (iii) melting of supersolid (SS) to NF in two steps via a DW/solid phase. 
Nature of the transition and an estimate of the corresponding critical temperature can be obtained by expanding the free energy $F$ w.r.t. the appropriate order parameter $\phi$  and rewriting it in the so called Landau-Ginzberg (LG) form,
\begin{equation}
F(\phi) = a + b\phi + c\phi^2 + d\phi^3 + e\phi^4 + \cdots 
\end{equation}     
where, the co-efficients $a,b,c,\cdots$ depend on the values of the parameters $t$, $\mu$ and $T$. 
In the homogeneous phase SF to NF transition can be captured from the vanishing of SF order parameter $\phi \equiv |\alpha|$. Near the transition the free energy $F$ can be expanded in even powers of $|\alpha|$ and the order parameter vanishes continuously at the critical temperature. The continuous transition from SF-NF and estimation of critical temperature from the LG free energy are discussed in Appendix\ \ref{AppendixA}. In the DW/solid phase the SF order parameter vanishes and density ordering takes the form $m_A = m_B \neq m_C$. 
For solid to NF transition the sub-lattice density difference $n_{A,B} -n_C = \delta$ plays the role of order parameter and the free energy contains both even and odd powers of $\delta$. 
Such MF free energy yields first order transition similar to three state Pott's model \cite{Prokofev05,Potts75} exhibiting a jump in density difference $\delta$ except at $\mu=3$ as also discussed in Appendix\ \ref{AppendixA}.
Next we focus on the melting of supersolid with increasing temperature. First, SS phase melts to a solid where the SF order parameter $(\alpha)$ vanishes continuously at the SS-solid phase boundary (see Fig.\ \ref{Egap_OP}a). Further increase of temperature leads to the melting of solid, and the sublattice density imbalance $(\delta)$ jumps to zero at the solid-NF boundary as depicted in Fig.\ \ref{Egap_OP}b. This is atypical of a first order transition. It can be noted from Fig.\ \ref{PD_MF}a, with increasing temperature the supersolid region shrinks and the two DW/solid lobes gradually merge to each other. Such a behavior of the supersolid phase is summarized in the $\mu$ vs $t$ phase diagram at different temperatures depicted in Fig.\ \ref{PD_MF}b.

%%%%%%%%%%%%%%%%%%%%%%%%%%%%%%%%%%%%%%%%%%%%%%%%%%%%%%%%%%%%%%%%%%%%%%%%%%%%%%%%%%%%%%%%%%%%
%%%%%%%%%%%%%%%%%%%%%%%%%%%%%%%%%%%%%%%%%%%%%%%%%%%%%%%%%%%%%%%%%%%%%%%%%%%%%%%%%%%%%%%%%%%%

At this point we would like to mention, although a first order transition from solid to NF phase with finite $T_c$ is observed at the MF level, the density ordering even at zero temperature can not survive at $\mu =3$ due to the effect of frustration. In absence of hopping $t=0$, this system becomes equivalent to an anti-ferromagnetic Ising model with vanishing critical temperature in absence of an external magnetic field which corresponds to $\mu=3$ in present case \cite{Houtappel50,Husimi50,Schick77}.
Such reduction of critical temperature at $\mu =3$ due to frustration can not be captured by simple MF method and requires more sophisticated techniques. 
This further motivates us to investigate the phase boundary particularly near $\mu=3$ by incorporating correlation via cluster mean field which we discuss in Sec.\ \ref{CMF_HCB}.

However, before going into that we complete the {\it mean field} analysis by analyzing the collective modes of these phases at finite temperature, as well determine the MF phase boundary from the energy gap vanishing phenomena in the next section.

\subsection{Collective modes of hardcore bosons}
\label{excitation_hcb}

At the MF level the collective excitation at zero temperature can be obtained by performing linear stability analysis of Gutzwiller wave function \cite{Sinha05,Saha14,Sinha07}.
Such method has also been extended for dissipative system involving the fluctuations of the density matrix \cite{Ray16}. Here we borrow the same methodology for density matrices \cite{Ray16} to obtain the finite temperature excitation spectrum of lattice bosons.
\begin{figure}[t]
\centering
\includegraphics[width=\columnwidth]{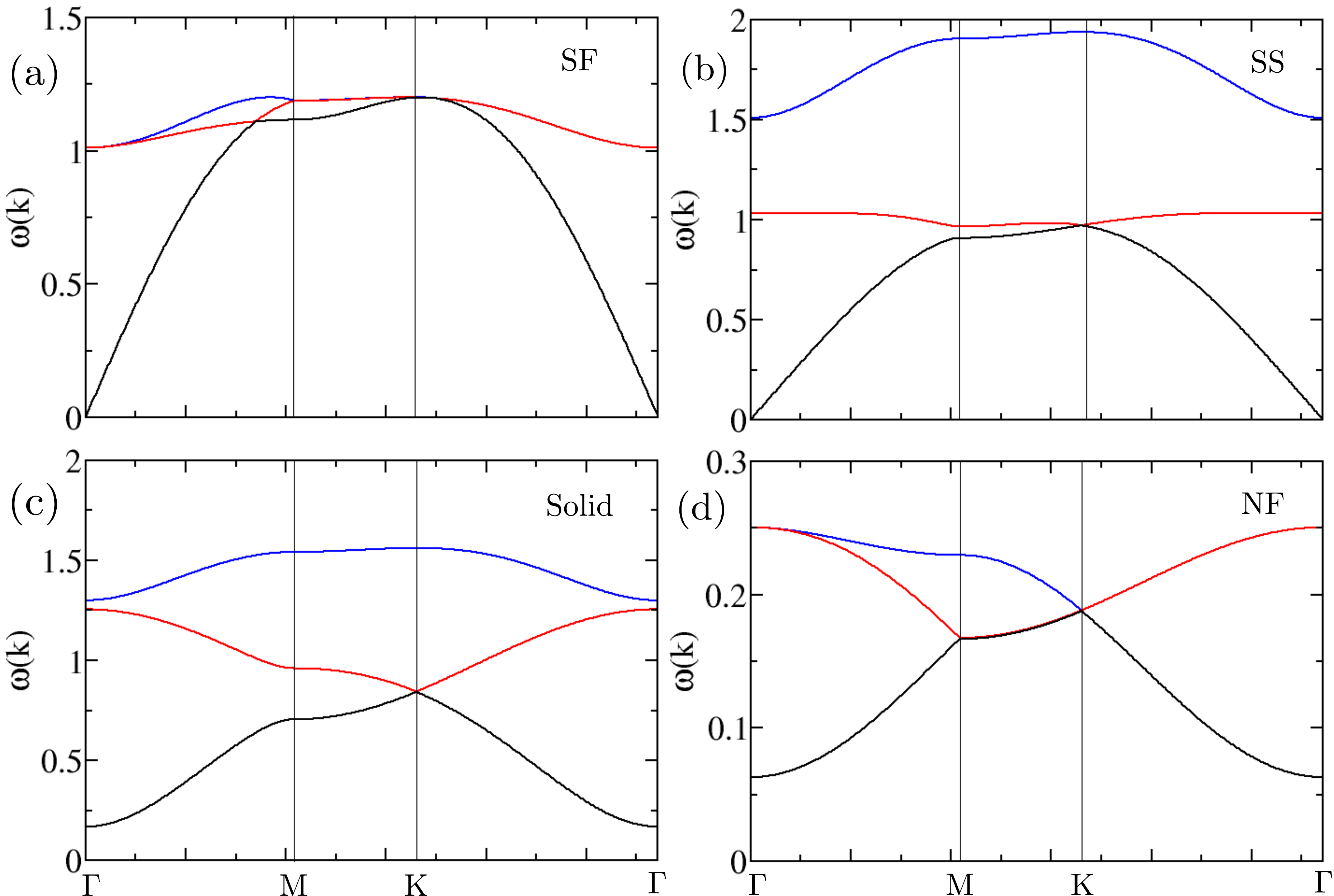}
\caption{(a) Collective excitation of SF phase $(\star)$ at temperature $T= 0.1$ and $\mu = 5.7$. and $t/V = 0.2$.
(b-d) Excitation spectrum of SS $(\bullet)$, Solid $(\square)$ and NF $(\triangle)$ phase at temperature $T/V=0.1$, $0.5$ and $0.9$ respectively, and $\mu = 3.5$. We choose $t/V = 0.2$ and the markers indicate the location of the points in the phase diagram in Fig.\ \ref{PD_MF}a.} 
\label{Excitation_hcb_T}
\end{figure}
Within the MF approximation, the time evolution of the density matrix $\hat{\rho}_i$ at $i$th site is governed by the following equation,
\begin{equation}
\dot{\hat{\rho}}_i = -i\left[\hat{H}_i^{\text{MF}},\hat{\rho}_i\right]
\label{den_evol}
\end{equation}  
where, the mean field Hamiltonian corresponding to the $i$th site is given by,
\begin{equation}
\hat{H}_i^{\text{MF}} = -t\left(\hat{a}_i^{\dagger} \bar{\alpha}_i + \hat{a}_i \bar{\alpha}_i^* \right) - \mu \hat{n}_i + V \hat{n}_i Q_i 
\label{H_MF}
\end{equation}
where, $\bar{\alpha}_i = \sum_{\langle i,j\rangle} \langle\hat{a}_j\rangle$ and $Q_i = \sum_{\langle i,j\rangle} \langle\hat{n}_j\rangle$. 
Linear stability analysis can be performed by introducing a small amplitude fluctuations around the steady state of the density matrix,
\begin{equation}
\rho_i^{a,b}(t) = \rho_{i,0}^{a,b} + \delta \rho_i^{a,b}(t)
\label{den_fluc}
\end{equation}
where, $\rho_{i,0}^{a,b}$ is the steady state value of the density matrix which satisfies $[\hat{H}_i^{\text{MF}},\hat{\rho}_{i,0}]=0$ and corresponds to the equilibrium distribution at temperature $T$, and $\delta \rho_i^{a,b}$ is the fluctuation around it at the $i$th site. Now, plugging Eq.\ \ref{den_fluc} into Eq.\ \ref{den_evol} we obtain,
\begin{equation}
\delta\dot{\hat{\rho}}_i = -i\left[\hat{H}_i^{\text{MF}},\delta\hat{\rho}_i\right] -i\left[\delta\hat{H}_i^{\text{MF}},\hat{\rho}_{i,0}\right]
\label{den_fluc1}
\end{equation}
where, $\delta\hat{H}_i^{\text{MF}}$ contains the fluctuation in mean-field terms of $\hat{H}_i^{\text{MF}}$. This is followed by substituting $\delta \rho_i^{a,b}(t) = \exp[i(\vec{k}.\vec{r}_i + \omega t)]\delta \rho_{\vec{k}}^{a,b}$, and retaining the terms linear in $\delta \rho_{\vec{k}}^{a,b}$ we obtain sets of linear equations describing the fluctuations in momentum space, and thereby construct the corresponding fluctuation matrix. It is important to mention that the fluctuations in the order parameters are induced by $\delta \rho_i(t)$ which is included in the linearized equations.  
The eigenvalues $\omega(k)$ of the fluctuations yields the dispersion of the collective excitations at finite temperatures. Also the stability of equilibrium density matrix is ensured by the condition ${\text{Im}[\omega]}=0$. 

\begin{figure}[t]
\centering
\includegraphics[width=\columnwidth]{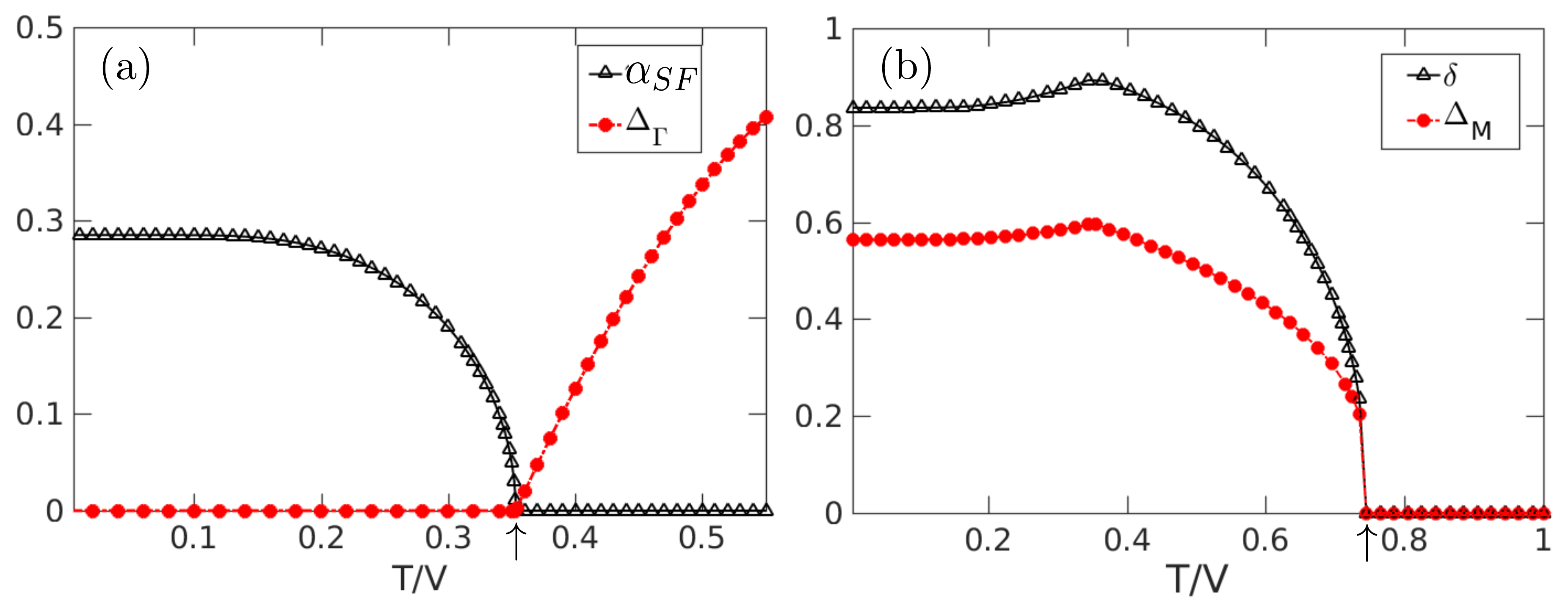}
\caption{(a) Energy gap opening at $\Gamma$ along with the vanishing of average SF order parameter $\alpha_{\mathrm{SF}}=\sum_i\alpha_i/3$ across SS-solid phase boundary.  
(b) Vanishing of sublattice density difference $\delta$ and energy gap $\Delta_M$ (at $M$ point) at solid-NF transition. We set the parameters $\mu/V = 3.5$ and $t/V = 0.2$. The transition temperatures are marked on the $T$ axis and its location in the phase diagram are marked by $(\uparrow)$ in Fig.\ \ref{PD_MF}a.}
\label{Egap_OP}
\end{figure} 

At a given temperature the steady state (equilibrium) order parameters $\alpha_i$'s and $m_i$'s are obtained by minimizing $F$, and by substituting them in Eq.\ \ref{den_fluc1} different branches of excitation spectra can be found for various phases as depicted in Fig.\ \ref{Excitation_hcb_T}. Note that these collective modes are plotted within one sub-lattice Brillouin zone of a triangular lattice (see the schematic in Fig.\ \ref{cmf_triangle}), where $\Gamma$, $M$ and $K$ represent the points $(k_x,k_y)$ as follows \cite{Owerre16,Berghmans18}:
\begin{equation}
\Gamma \equiv (0,0), \quad M \equiv \left(\frac{2\pi}{3},0\right), \quad K \equiv \left(\frac{2\pi}{3},\frac{2\pi}{3\sqrt{3}}\right)
\end{equation}  
From the characteristic features of the excitation modes we identify different symmetry broken and unbroken phases and transition between them. 
Both SF and SS phases are characterized by the gapless sound modes $\omega \sim c_s|k|$ for $|k| \ll 1$ due to the presence of SF order parameter (see Fig.\ \ref{Excitation_hcb_T}a,b).
On the other hand a gap opens up for both the insulating phases {\it i.e} DW and NF as shown in Fig.\ \ref{Excitation_hcb_T}(c,d). Therefore both the transition SF-NF and SS-solid at finite temperature can be identified from the energy gap opening at $|k|=0$ ($\Gamma$ point). As shown in Fig.\ \ref{Egap_OP}a the SF order parameter vanishes at the critical temperature above which the energy gap increases continuously.  It can be noted that in the homogeneous phases (SF and NF) there can be only one mode due to one sub-lattice structure, however, within the reduced Brillouin zone (BZ) we obtain three excitation branches out of which two modes become gapless along $M-K-\Gamma$ line due to sublattice symmetry \cite{Murthy97,Berghmans18}.
In translation symmetry broken phases (solid and SS) such a degeneracy is lifted except at $K$ point, and thus they can be identified from the energy gap say, $\Delta_M$ at the $M$ point of BZ for the branches which are degenerate at $K$ point. For solid to NF transition the variation of energy gap of these two modes at $M$ point and density difference $n_{A,B} -n_C$ with increasing temperature are shown in Fig.\ \ref{Egap_OP}b. Both the quantities undergo a sharp jump at the critical temperature and vanishes in the homogeneous NF phase as a consequence of first order transition. We also point out that close to zero temperature the excitation spectrum obtained in this method are in agreement with those obtained from spin wave analysis \cite{Murthy97}. 
Although the exact nature and critical behavior of the transition is beyond the scope of MF analysis, however, different phases and transition between them at finite temperatures can be identified from the above mentioned features of the collective modes and can be relevant for experimental detection.

In the next subsection we consider cluster of the unit cells in order to incorporate the effect of correlation in a systematic way and also discuss how it improves the phase diagram particularly near $\mu=3$ where the effect of frustration is much more pronounced and cannot be captured from simple MF theory.

\subsection{Cluster mean field theory}
\label{CMF_HCB}

\begin{figure}[t]
\centering
\includegraphics[width=\columnwidth]{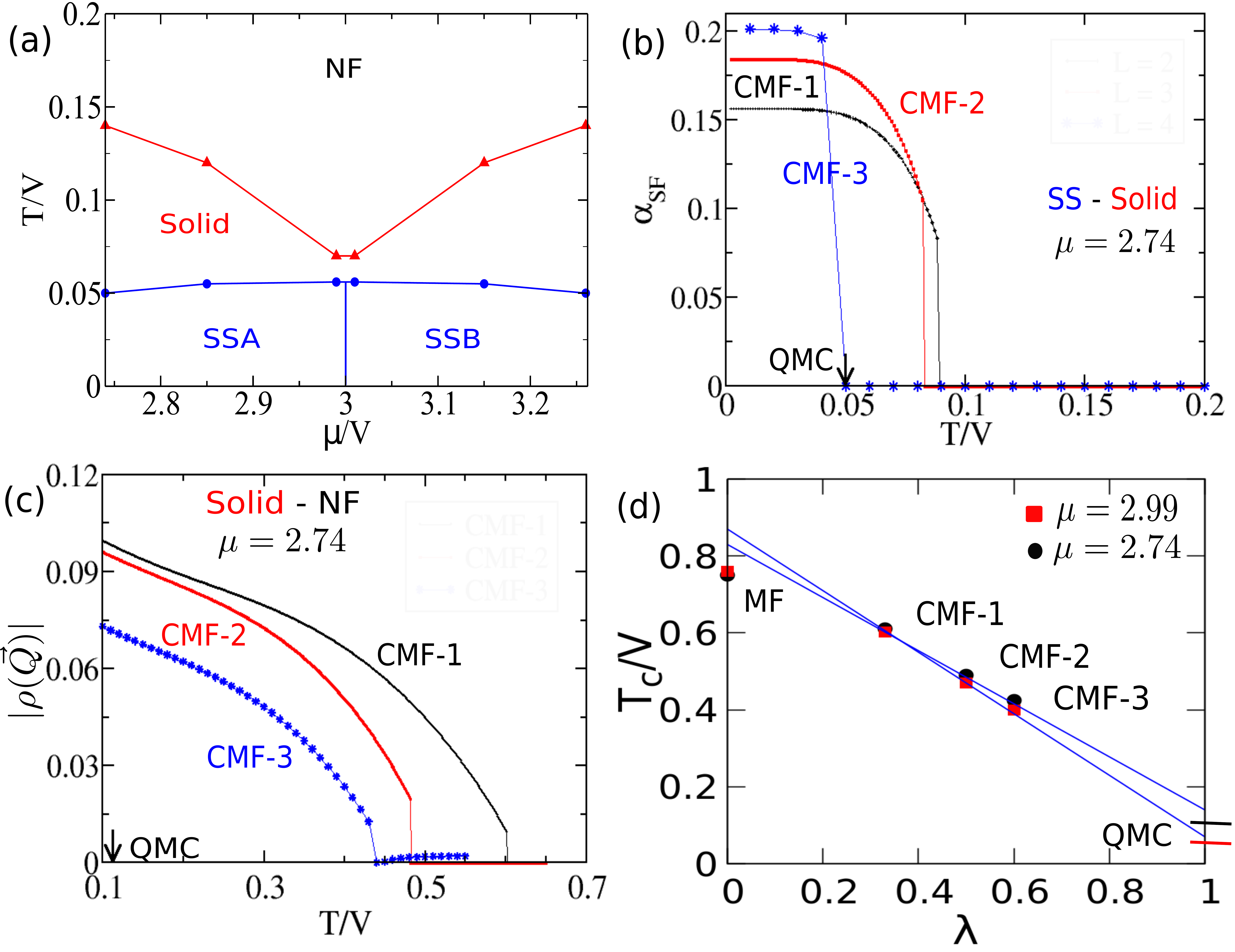}
\caption{{\it Finite temperature phase diagram of hardcore bosons in triangular lattice using CMF:} (a) in $\mu$ vs $T$ plane for $t=0.1$. (b-c) Vanishing of SF order parameter $\alpha_{\text{SF}}$ and $\rho(\vec{Q})$ in SS to Solid and Solid to NF transition respectively for different cluster sizes mentioned therein. (d) Infinite cluster size extrapolation of the Solid to NF transition temperature ($T_c$) obtained from the vanishing of $\rho(\vec{Q})$. The extrapolated values are used to obtain the phase diagram in (a) and the phase boundary agrees well with the QMC results \cite{Prokofev05}.}
\label{PD_CMF}
\end{figure}

We use the CMF method discussed in Sec.\ \ref{model} and investigate the transition temperature between the phases, particularly near $\mu=3$. Different phases are characterized as follows. The presence of superfluidity is determined by the non-vanishing SF order parameter, $\alpha_{\text{SF}} = \sum_{i=1}^{N}\langle \hat{a}_i\rangle/N$, $N$ being the number of lattice sites within the cluster. The density ordering with two sublattice structure in DW/solid and in SS phase is characterized by,
\begin{equation}
\rho(\vec{Q}) = \frac{1}{N} \sum_{i=1}^{N} ~ \langle \hat{n}_i \rangle ~ e^{i\vec{Q}.\vec{r}_i}, \quad \vec{Q} = \left( 4\pi/3,0 \right)
\end{equation}
As a result of two step melting of supersolid with increasing temperature, first $\alpha_{SF}$ vanishes indicating the SS-solid transition, followed by the vanishing of $|\rho(\vec{Q})|$ at higher temperature showing solid-NF transition (see Fig.\ \ref{PD_CMF}a). We observe that the critical temperature particularly for solid-NF transition varies significantly with the cluster size, and becomes more and more accurate with its increasing size, as depicted in Fig.\ \ref{PD_CMF}(b,c). 
Further, in order to improve the phase boundary we have performed a finite cluster-size scaling \cite{Yamamoto12} by analyzing the solid-NF transition temperature $T_c$ as a function of a scaling parameter $\lambda = \frac{N_B}{Nz/2}$, where, $N_B$ is the number of bonds in a cluster and $z$ is the coordination number which is $6$ for triangular lattice. The values of $T_c$ obtained from different cluster sizes are plotted as a function of $\lambda$ in Fig.\ \ref{PD_CMF}d. The data are fitted by a straight line and extrapolated to the thermodynamic limit $(\lambda \rightarrow 1)$ to extract $T_c$ more accurately. The transition temperature obtained from the Quantum Monte Carlo (QMC) study \cite{Prokofev05} is marked by the horizontal cuts. As an example, for $\mu=2.74$ we obtain the extrapolated value of $T_c/V$ is $0.14$, which is fairly close to the exact value $T_c/V=0.1033$; similarly, for the same $\mu=2.74$ the SS-solid transition temperature obtained using larger cluster is $0.049$ which agrees closely with the QMC data $T_c/V=0.05$ \cite{Prokofev05}. Note that this scaling analysis is different from finite size scaling which is typically done in the numerical analysis of finite size systems. The resulting phase boundary in the $\mu$ vs $T$ plane obtained in this way is shown in Fig.\ \ref{PD_CMF}a, which is in a very close agreement with the QMC results \cite{Prokofev05}. Thus our analysis presents how the effect of correlation can be incorporated with increasing order of cluster size resulting in a remarkable improvement of the phase boundary near $\mu=3$. However, we do not focus on the type of SS phase formed at $\mu=3$ and the transition between the two types of SS phases namely, SSA and SSB (as mentioned in Sec.\ \ref{MF_HCB}) \cite{Damle05,Prokofev05,Yamamoto12} which is beyond the scope of the present study.
   
\section{Phases and collective excitation of boson with finite $U$}
\label{scb}

In a more realistic scenario concerning the experiment, in this section we discuss the phases and collective modes of bosons with finite onsite interaction $U$. Our aim is to study the new phases that appear because of finite $U$, their melting with increasing temperature and to characterize these phases from the collective excitation at low temperature.

\subsection{Zero and finite temperature phases}

\begin{figure}[t]
\centering
\includegraphics[clip=true, width=1\columnwidth]{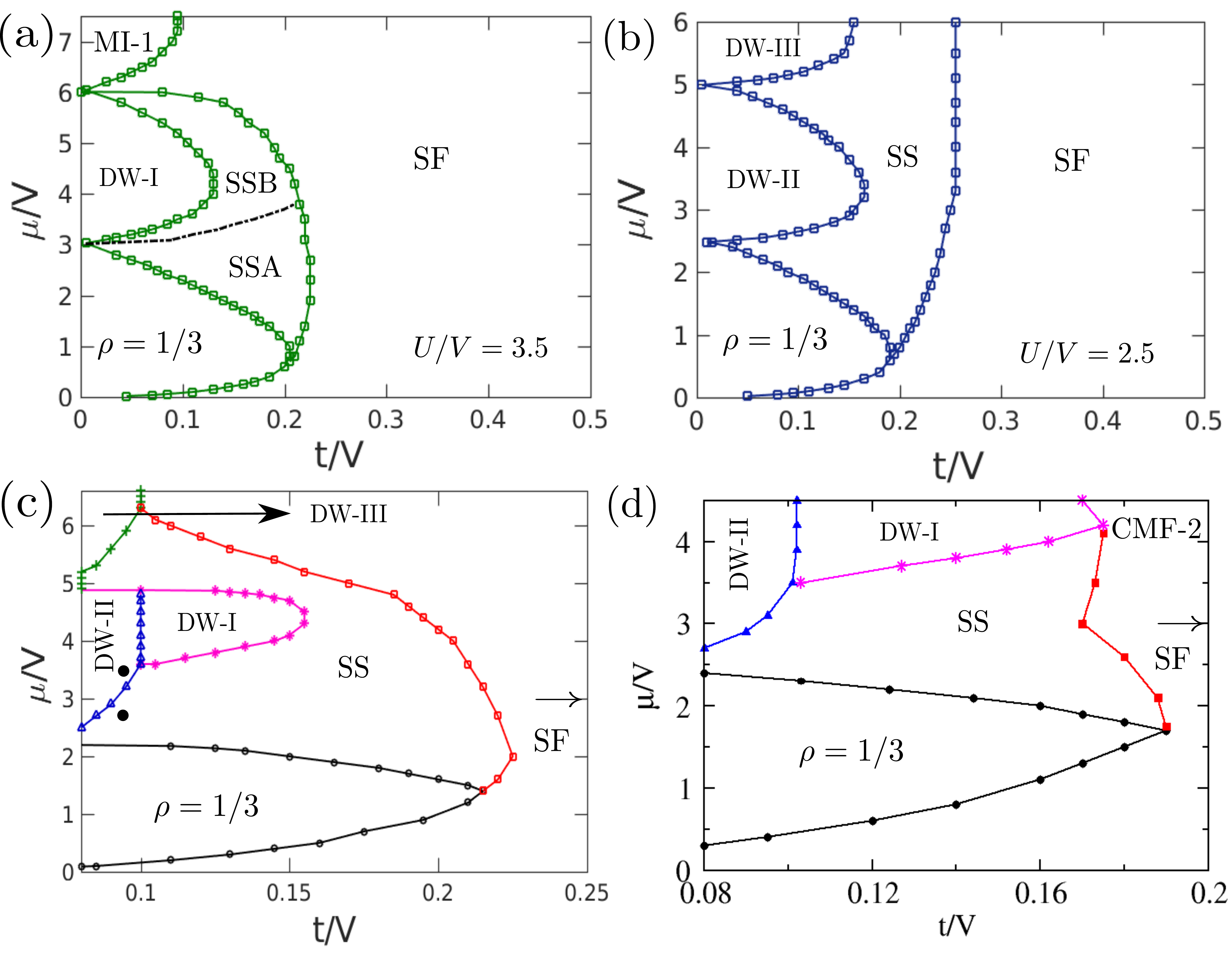}
\caption{{\it Zero temperature phase diagram of bosons in triangular lattice:} (a-b) in $\mu$ vs $t$ plane for different values of $U$ mentioned therein, (c-d) for varying onsite interaction $U=30t$ obtained by using MF and CMF-2 respectively. Details of these phase diagrams are discussed in the text.}
\label{PD_SCB}
\end{figure}

To this end we consider the Gutzwiller variational wave function for three sub-lattices $i=A$, $B$ and $C$ which constitutes the unit cell of a triangular lattice, is given by,
\begin{equation}
\vert\Psi\rangle = \prod_i \vert \psi_i \rangle, \quad \vert\psi_i\rangle = \sum_n f_{i}^n \vert n\rangle_i
\end{equation}  
where, $\vert n\rangle_i$ represents the Fock state with occupation $n$ and probability amplitude $|f_n^{i}|^2$ at the $i$th site. For the MF calculation we truncate the Fock Hilbert space suitably and the normalization is set such that $\sum_n |f_n^i|^2=1$. It is also ensured that the truncation is sufficient to capture the phases which we discuss in the following. We numerically minimize the energy functional $E=\langle \Psi \vert \hat{H}_{MF} \vert \Psi \rangle$ and chart out the phases as a function of $\mu$ and $t$ for different values of onsite interaction $U$, illustrated in Fig.\ \ref{PD_SCB}. 
Since the particle-hole symmetry between $\rho=1/3$ $(1,0,0)$ and DW-I $(1,1,0)$ is destroyed, these insulating lobes are no longer symmetric as observed in Fig.\ \ref{PD_SCB}a. Lowering $U<3V$, DW phases of higher filling $(n_0,0,0)$, with $n_0=2,3,~\cdots$ appears; as shown in Fig.\ \ref{PD_SCB}(b,c). Also DW lobe with $\rho=2/3$ changes from DW-I to DW-II $(2,0,0)$ with decreasing $U$. The continuous deformation of the insulating lobes with decreasing $U/V$ is shown in Appendix\ \ref{AppendixC}. In Fig.\ \ref{PD_SCB}c we have presented a phase diagram in $\mu-t$ plane with varying $U$ which scales with respect to $t$. It can be noted that this simple mean field theory captures all the phases as observed in more exact QMC studies \cite{Gan07}, {\it albeit} with an expected difference in the phase boundary and an extended SS region over the DW phases. To improve the phase boundary we incorporate correlation via considering clusters using the zero temperature CMF method described in Sec.\ \ref{model}, and the resulting phase diagram is shown in Fig.\ \ref{PD_SCB}d. 

%%%%%%%%%%%%%%%%%%%%%%%%%%%%%%%%%%%%%%%%%%%%%%%%%%%%%%%%%%%%%%%%%%%%%%%%%%%%%%%%%%%%%%%%%%%%
%%%%%%%%%%%%%%%%%%%%%%%%%%%%%%%%%%%%%%%%%%%%%%%%%%%%%%%%%%%%%%%%%%%%%%%%%%%%%%%%%%%%%%%%%%%%

Next we focus on the melting of $\rho=2/3$ lobe as temperature is increased at a fixed chemical potential $\mu=3$ (marked by `$\rightarrow$' in Fig.\ \ref{PD_SCB}c,d), particularly, the new phases such as DW-II and the SS around it which occur between $0.05 \lesssim t/V \lesssim 0.12$ at zero temperature. With increasing $T$, SS phase goes through a two step melting to normal fluid via a solid phase as depicted in Fig.\ \ref{PD_MF_T}a,b. As $t/V$ is increased solid region shrinks and thereby the gap between critical temperatures corresponding to SS-solid and solid-NF transition decreases, and finally vanishes at the SF-NF boundary. On the other hand, as $t/V$ is decreased the effect of $U$ starts playing an important role in the melting of DW-II to NF, and the transition temperature increases with increasing $V/U$ as observed from both MF and CMF analysis (see Fig.\ \ref{PD_MF_T}a,b). However, as expected with decreasing $t/V$ i.e. in a more correlated regime the difference between MF and CMF results of such transition are more pronounced. It is also important to mention that SS-SF boundary (vertical line in Fig.\ \ref{PD_MF_T}a,b) is not affected by the thermal fluctuation, because SF order always vanishes earlier than the DW order with increasing $T$, thereby discards the possibility to observe SS-SF transition due to temperature. 

\subsection{Collective excitations of soft core bosons}
In this section we discuss collective excitation, particularly of the new phases and investigate their transitions at finite temperature. 
To understand the appearance of new stable phases by varying the onsite interaction $U$, we first consider the excitations of insulating phases 
at zero temperature and for vanishing hopping strength. For sufficiently large onsite repulsion $U$ similar density ordering like HCB occurs and DW phases can be classified as $(n_0,n_0-1,n_0-1)$ and $(n_0,n_0,n_0-1)$ representing number of particles at sublattice (A,B,C) respectively. First $(1,0,0)$ DW phase is appeared, which has two degenerate particle excitations at A, B sites with energy $E^{A,B}_{p}=3V-\mu$ and particle (hole) excitation energy $E^{C}_{p}= U -\mu$ ($E^{C}_{h}=\mu$) at site C. The hole excitation $E^{C}_{h}$ becomes unstable at $\mu=0$ and for large $U$ the instability of particle excitation $E^{A}_{p}$ at $\mu=3V$ leads to the formation of DW-I with filling $2/3$. 
%%%%%%%%%%%%%%%%%%%%%%%%%%%%%%%%%%%%%%%%%%%%%%%%%%%%%%%%%%%%%%%%%%%%%%%%%%%%%%%%%%%%%%%%%%%%
\begin{figure}[t]
\centering
\includegraphics[clip=true, width=1\columnwidth]{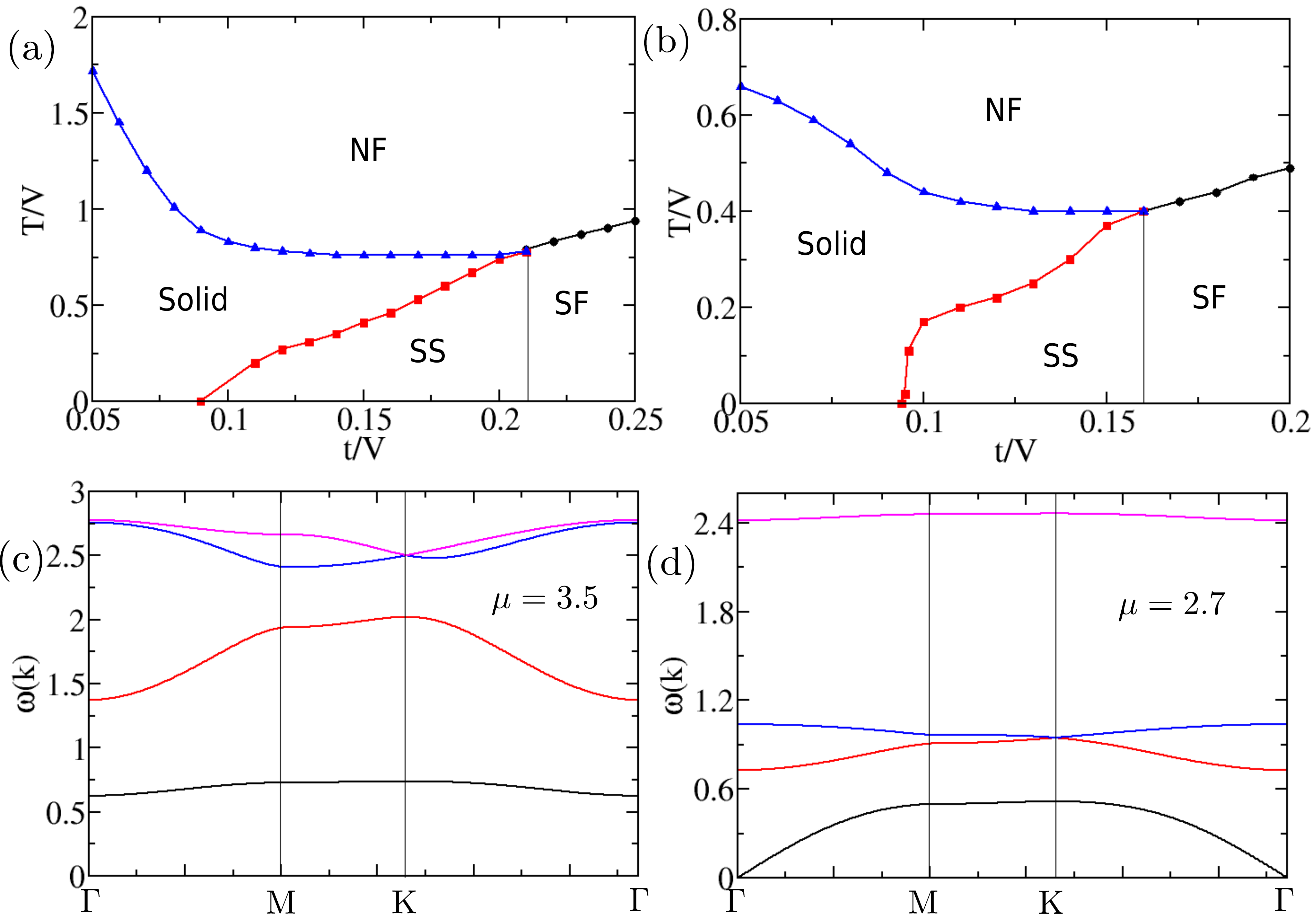}
\caption{{\it Finite temperature phase diagram of bosons in triangular lattice:} (a-b) in $t/V$ vs $T/V$ plane for $U=30t$ and $\mu=3$ (marked by `$\rightarrow$' on secondary y-axis in Fig.\ \ref{PD_SCB}c,d), using MF and CMF-2 respectively. (c-d) Collective excitation of DW-II and SS around it, for $\mu=3.5$ and $\mu=2.7$ respectively as marked by ($\bullet$) in Fig.\ \ref{PD_SCB}c, at very low temperature $T=0.01$.}
\label{PD_MF_T}
\end{figure}
%%%%%%%%%%%%%%%%%%%%%%%%%%%%%%%%%%%%%%%%%%%%%%%%%%%%%%%%%%%%%%%%%%%%%%%%%%%%%%%%%%%%%%%%%%%%
However, this scenario changes for $U<3V$ when the particle excitation $E^{C}_{p}$ at C becomes unstable first at $\mu = U<3V$ and a new DW-II phase appears. In Fig.\ \ref{PD_SCB}b note that these are boundary points of $\rho=1/3$ phase in the atomic limit $(t=0)$. In DW-I the low lying modes are degenerate hole excitations $E^{A,B}_{h}=\mu -3V$ and particle excitation $E^{C}_{p}=6V-\mu$ which become unstable at $\mu = 3V$ and $6V$ respectively. For large $U$ these low lying modes are similar to those of HCB and we skip that discussion. Instead we focus on $U<3V$ regime and the DW-II phase which has degenerate particle excitation $E^{A,B}_{p}=6V -\mu$ and two lower energy particle (hole) excitation $E^{C}_{p}=2U -\mu$ $(E^{C}_{h}=\mu-U)$. These determine the stability of this phase within the region $U<\mu<2U$ (see that these are the boundaries of DW-II for $t=0$ in Fig.\ \ref{PD_SCB}b). For finite $t$, its low lying excitations are shown in Fig.\ \ref{PD_MF_T}c. As $t$ is increased it melts to SS phase along with the vanishing of energy gap at $\Gamma$ point. The low lying excitations of the SS surrounding DW-II are shown in Fig.\ \ref{PD_MF_T}d. Comparing low energy excitation of the two types of SS formed around DW-I and DW-II phases, we see that although both of them show gapless sound mode, however, a gap between lowest two branches of excitation at $K$ point exists for $U<3V$, whereas, it vanishes for $U>3V$ which is the reminiscence of lowest degenerate hole excitation of DW $(1,1,0)$ (see Fig.\ \ref{Excitation_hcb_T}b). This completes our analysis on the characteristic features of the excitation of new phases at finite $U$ and low temperature.  

\section{Summary}
\label{conclu}

To summarize, we studied various phases of bosons and their collective excitations in a triangular lattice at finite temperature both for onsite hardcore repulsion $(U \rightarrow \infty)$ and for finite $U$. The effect of lattice geometric frustration and strong correlation between the atoms play a crucial role in the formation of different phases and transition between them. 
We obtained finite temperature phase diagram using single site mean field method approximating the density matrix as a product of single site density matrices and thereby ignoring the inter-site correlation. The main advantage of doing MF is to obtain a semi-analytic estimate of the phase boundaries within the framework of Landau theory, particularly for hardcore bosons. We also performed more accurate cluster mean field theory and compare both the results in order to gain information about the effect of correlation. Moreover, the collective excitation frequencies of the various phases at finite temperature are calculated from the time dependent fluctuations of the density matrix.  Different characteristics of such low lying excitations carry the signature of various phases and signals the transition between them, which can be used for their experimental detection. 

As a result of the interplay between lattice frustration and nearest neighbor repulsion between the atoms a stable supersolid phase is formed around the DW phases with different filling (depending on $\mu$ and $U$). With increasing temperature two step melting of the SS phase is observed; first, SF order parameter vanishes at much lower temperature which scales with hopping strength and then the solid phase melts to homogeneous NF at higher temperature comparable with nearest neighbor interaction. As $t/V$ increases, the gap between these two transition temperatures eventually vanishes and merges to SF-NF phase boundary. Within MF theory we observe a continuous transition between SS-DW/solid and SF-NF phases. The gapless sound mode at $k=0$ present in SF and SS phases becomes gaped at their respective phase boundaries. Whereas, solid phase undergoes a first order transition to NF with increasing $T$, characterized by a jump in DW order parameters such as sublattice density imbalance as well as $\rho(\vec{Q})$ at $\vec{Q}=(4\pi/3,0)$. The degenerate excitation modes of the DW phase at $K$ point also becomes gapless at $M$ point of the Brillouin zone during the transition to homogeneous NF. The behavior of the collective excitation of different phases at finite $T$ are important in the context of recent cold atom experiments where the low energy Goldstone and Higgs modes in a supersolid are detected using spectroscopic measurement \cite{Esslinger17} or using the time-of-flight experiments \cite{Ferlaino19,Pfau19,Stringari19}.

Although simple MF theory provides the qualitative understanding of the finite temperature phases, however, as expected it fails to capture the exact nature of transition as well as the quantitative estimate of transition temperature, particularly the melting of SS and solid phases due to the effect of frustration. For hardcore bosons at and around $\mu=3$, enhanced effect of frustration significantly reduces the melting temperature of DW phase. For $t=0$ the system of hardcore bosons becomes equivalent to disordered anti-ferromagnet with vanishing critical temperature. By incorporating finite cluster-size scaling, CMFT can successfully capture such reduction of melting temperature around $\mu=3$ which is also in agreement with the QMC results \cite{Prokofev05,Gan07}. This indicates both the effect of correlation and frustration can be captured by CMFT which is thus an useful tool to study the finite temperature phases of interacting bosons in an optical lattice and phase transitions between them. Such a method can further be extended to study the non-equilibrium dynamics of strongly correlated lattice bosons at finite temperature.

In conclusion, we have investigated the effect of correlation arising from lattice frustration and interaction systematically using {\it mean field} as well as {\it cluster mean field} theory, and identified the different phases from their characteristic low lying excitation. These collective modes and their behavior at finite temperature which we discussed can be probed experimentally using the similar line of thought as in recent cold atom setups \cite{Esslinger17,Ferlaino19,Pfau19,Stringari19}.  

\section*{Acknowledgement}
SR acknowledges financial support from the Israel Science Foundation (Grant No. 283/18).

\appendix

\section{Landau-Ginzberg theory of phase transition}
\label{AppendixA}

In the following subsections, we will discuss the nature of the transitions between different phases as well as the transition temperature predicted from LG theory as a result of the variation of temperature.

\subsection{Superfluid to normal fluid transition}
\label{AppendixA1}

\begin{figure}[t]
\centering
\includegraphics[width=\columnwidth]{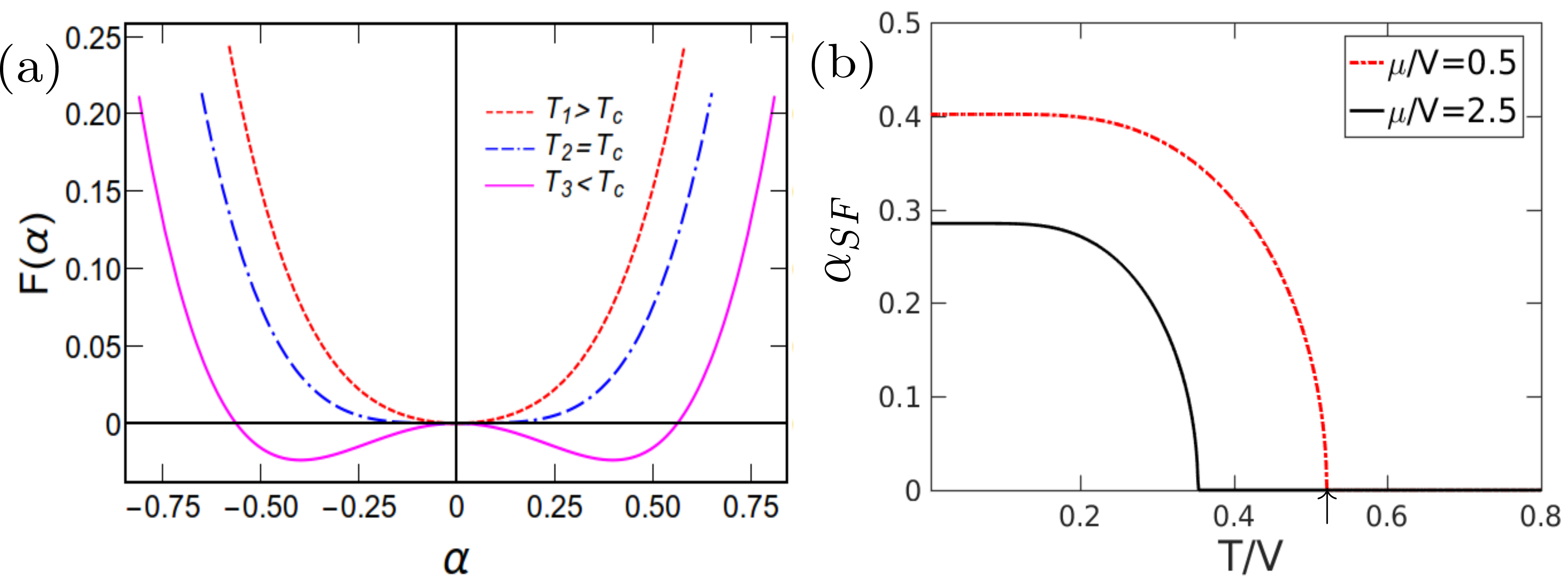}
\caption{(a) Landau free energy $F$ as a function of the average SF order parameter $\alpha_{\mathrm{SF}} = \alpha$ for different T, each of them is scaled by subtracting $F(\alpha=0)$. Parameters are $t/V=0.2$ and $\mu/V=0.5$. (b) $\alpha_{\mathrm{SF}}$ with increasing temperature at $t/V=0.2$ for different $\mu/V$ as mentioned in the inset. The numerically obtained SF-NF transition temperature $T_c$ agrees with that from the LG theory as marked by $(\uparrow)$ on the $T/V$ axis.}
\label{AppenA_Fig2}
\end{figure}

In the homogeneous SF phase, average free energy of the system given in Eq.\ \ref{free_en} can be written as,
\begin{equation}
F = -6t\alpha^2 -\frac{\mu}{2}(1+m) +\frac{3V}{4}(1+m)^2 +T\sum_{\sigma=+,-} \lambda_{\sigma}\log \lambda_{\sigma}
\end{equation} 
where, $\lambda_{\pm} = (1\pm \sqrt{m^2+4\alpha^2})/2$. We have assumed $\alpha$ to be a real parameter without any loss of generality and because of homogeneity of the superfluid we have put $\alpha_i = \alpha$ and $m_i = m$. 
Now the free energy can be expanded in a power series of SF order parameter $\alpha$ as follows,
\begin{equation}
F = a(\mu,t,m,T) + b(\mu,t,m,T)\alpha^2 + c(\mu,t,m,T)\alpha^4 + \cdots
\label{F_gamma}
\end{equation}
This is the Landau-Ginzburg form of second order phase transition. Thus the critical temperature can be estimated by numerically finding the values of $m$ and then evaluating the co-efficients $a, b, c$ of Eq.\ \ref{F_gamma}. In Fig.\ \ref{AppenA_Fig2}a we have plotted $F$ at $T<T_c$, $T=T_c$ and $T>T_c$. 

As mentioned in the main text, we observe with increasing temperature SF order parameter vanishes continuously at the SF-NF phase boundary. In Fig.\ \ref{AppenA_Fig2}b we have shown the variation of SF order parameter as a function of temperature. It can be noted that the numerically obtained value of the critical temperature (as marked by $\uparrow$ in Fig.\ \ref{AppenA_Fig2}b) agrees with that estimated from the LG theory. 

\subsection{Solid to normal fluid transition}
\label{AppendixA2}

Density wave phase is characterized by vanishing of SF order parameters $(\alpha=0)$ and non-zero value of density order $(m_i \neq m_{\bar{i}})$. Let us consider, $n_A = n_B = (n-\delta)$ and $n_C = (n+2\delta)$, where $\delta$ is the DW order parameter. Following this parametrization, free energy in terms of $m$ and $\delta$ is given by,
\begin{eqnarray}
F &=& \frac{3V}{4}(m^2+2m-4\delta^2+1) - \frac{\mu}{2}(1+m) \nonumber \\
&+& \frac{T}{3} \sum_{\substack{i=A,B,C \\ j=1,2}} \lambda_i^j \log \lambda_i^j
\end{eqnarray} 
where, $\lambda_{A}^{1,2} = \lambda_{B}^{1,2} = [1\pm (m-2\delta)]/2$ and $\lambda_{C}^{1,2} = [1\pm (m+4\delta)]/2$. 
For a given value of $m$, we can write the free energy in a power series of $\delta$ which is given by,
\begin{eqnarray}
F &=& a(\mu,t,m,T) + b(\mu,t,m,T)\delta + c(\mu,t,m,T)\delta^2 \nonumber \\
&+& d(\mu,t,m,T)\delta^3 + \cdots
\end{eqnarray}
Non-zero values of the co-efficients $b$ and $d$ imply that, this is the Landau-Ginzburg form of first order phase transition. Critical temperature can be obtained by evaluating the co-efficients $a, b, c, d$ by numerically finding the value of $m$. 

\begin{figure}[t]
\centering
\includegraphics[width=\columnwidth]{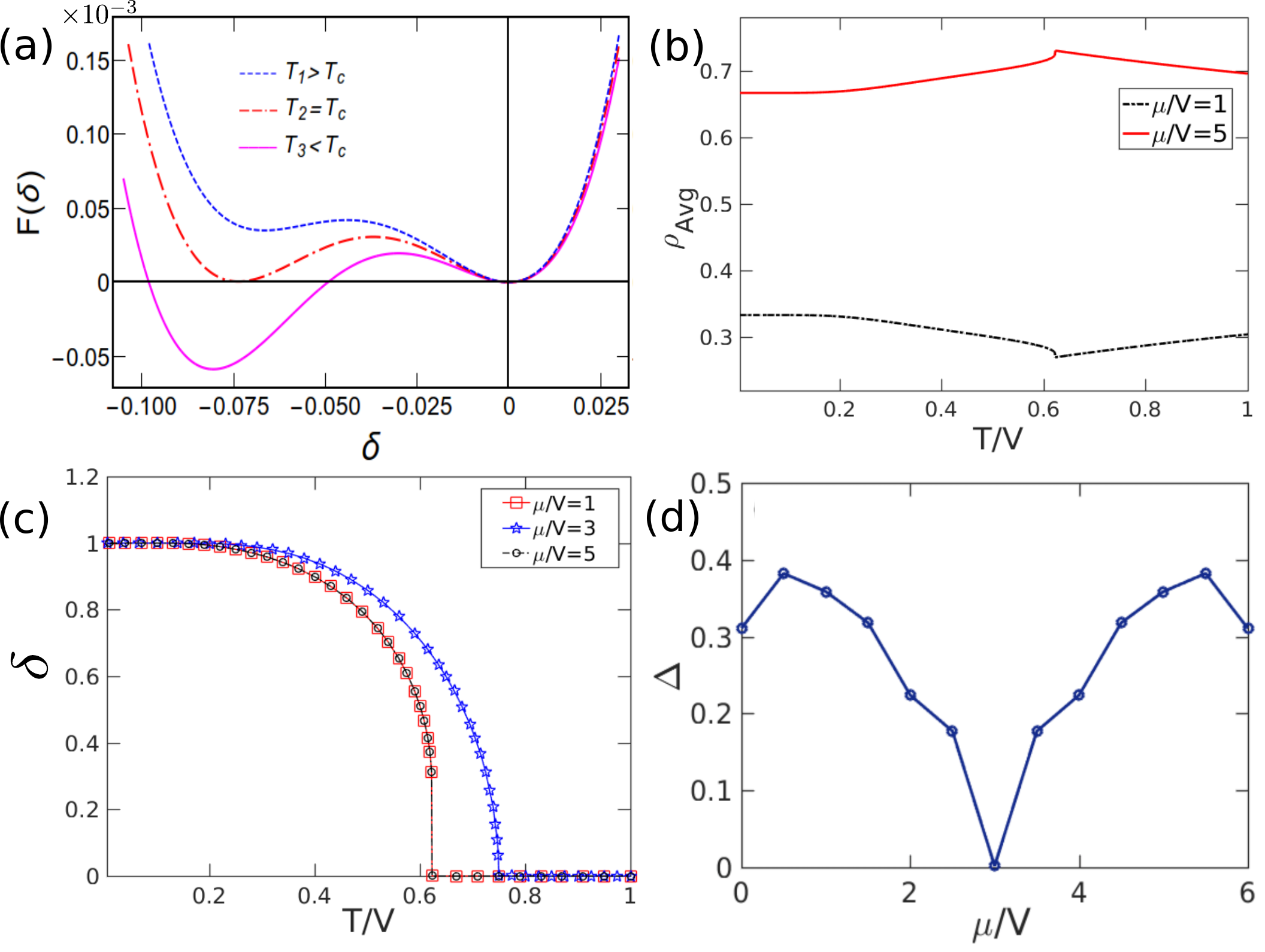}
\caption{(a) Landau free energy $F$ is plotted as a function of $\delta$, scaled by subtracting $F(\delta=0)$ for $\mu/V=5$ and for different temperatures mentioned therein. (b-c) Average density $\rho_{avg}$ and sublattice density imbalance $\delta$ with increasing $T$ for different $\mu/V$. (d) Magnitude of the jump $\Delta$ in $\delta$ at $T_c$ for solid-NF transition vs $\mu/V$. We set $t/V = 0$.}
\label{AppenA_Fig3}
\end{figure}

In Fig.\ \ref{AppenA_Fig3}a we have shown the typical variation of LG free energy as a function of $\delta$. At the critical $T_c$, $\delta$ corresponding to the minima of $F$ exhibits a jump from a finite value to zero as depicted in Fig.\ \ref{AppenA_Fig3}c. Such a jump is a characteristic feature of first order transition from solid to NF phase. However, the magnitude of jump reduces as $\mu/V$ becomes closer to 3 and vanishes at $\mu=3$ as shown in Fig.\ \ref{AppenA_Fig3}d. We would like to point out that a similar phenomena was observed in case of SS to SF phase transition at zero temperature \cite{Yamamoto12}. Both the average density $\rho_{avg}$, the sublattice density imbalance $\delta$ and its jump $\Delta$ at solid-NF phase boundary behaves symmetrically away from $\mu/V=3$ as can be noted from Fig.\ \ref{AppenA_Fig3}b,c,d respectively. 

\section{Zero temperature phase diagram of softcore bosons}
\label{AppendixC}

\begin{figure}[ht]
\centering
\includegraphics[width=\columnwidth]{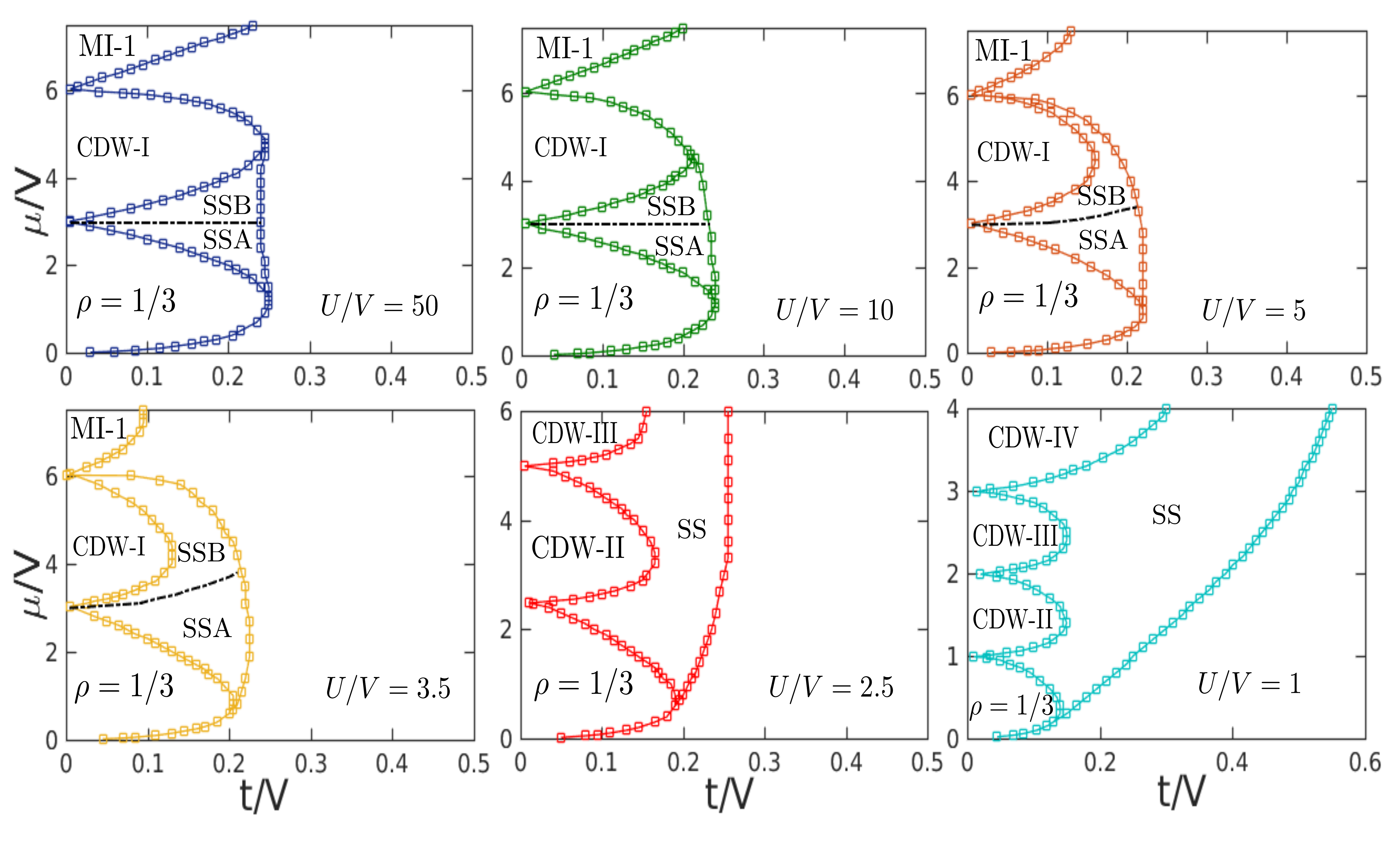}
\caption{Zero temperature mean field phase diagram of bosons with finite $U$ in the $\mu-t$ plane. Different values of $U$ are mentioned in the figure inset.}
\label{AppenC_Fig1}
\end{figure}

Here we show how the breaking of particle hole symmetry due to finite $U$ deforms the insulating lobes and new density wave phases with higher filling appear. In Fig.\ \ref{AppenC_Fig1} we have demonstrated this issue, where $U$ is gradually decreased from a large value for which we recover the hardcore boson phase diagram. With decreasing $U$ these lobes deforms and below $U=3V$, $\rho=2/3$ phase changes from $(1,1,0)$ to $(2,0,0)$ along with the appearance of other DW phases with higher filling, and an extended supersolid forms above these insulating phases.

\end{document}